\documentclass[letterpaper,aps,floatfix,twocolumn,showpacs]{revtex4}
\usepackage{graphicx}
\usepackage{epsfig}     
\usepackage{amsmath}
\usepackage{amssymb}
\usepackage{hyperref}
\usepackage{alltt}
\usepackage{setspace}
\newcommand{\old}[1]{{}}

\begin{document}

\title{On the physical inadmissibility of ILES for simulations of Euler
equation turbulence}

\author{J. Glimm}
\email{glimm@ams.sunysb.edu}
\affiliation{Stony Brook University, Stony Brook NY 11794}
\author{Baolian Cheng}
\email{bcheng@lanl.gov}
\affiliation{Los Alamos National Laboratory, Los Alamos NM 87545}
\author{David H. Sharp}
\email{dcso@lanl.gov}
\affiliation{Los Alamos National Laboratory, Los Alamos NM 87545}
\author{Tulin Kaman}
\email{tkaman@uark.edu}
\affiliation{University of Arkansas, Fayetteville AR 72701}

\date{\today}

\begin{abstract}
We present two main results. The first is a plausible validation argument for
the principle of a maximal rate of entropy production for
Euler equation turbulence. This principle can be seen as an 
extension of the second law of thermodynamics. In our 
second main result, 
we examine competing models for large eddy simulations of 
Euler equation (fully developed) turbulence. We compare schemes with no subgrid
modeling, implicit
large eddy simulation (ILES) with limited subgrid modeling
and those using dynamic subgrid scale
models. Our analysis is based upon three fundamental physical principles:
conservation of energy, the maximum entropy production rate and
the principle of universality for multifractal
clustering of intermittency. 
We draw the
conclusion that the absence of subgrid modeling, or its partial
inclusion in ILES solution violates the maximum entropy dissipation rate
admissibility criteria. We identify circumstances in which the
resulting errors have a minor effect on specific observable quantities and
situations where the effect is major.

Application to numerical modeling of the deflagration to detonation
transition in type Ia supernova is discussed.
\end{abstract}
\date{}
\maketitle


\section{Introduction}
\label{sec:intro}

The solutions of the Euler equation for fluid dynamics are not unique.
An additional law of physics, in the form of an entropy principle, is
needed to ensure a physically meaningful solution. Wild and manifestly
nonphysical solutions have been studied extensively \cite{DelSze09,DelSze10} and
offer counter examples to studies of the Euler equation as a model for
fully developed turbulence. This paper is concerned with the nonuniqueness
for Euler equation solutions that are the limit of Navier-Stokes solutions as
the viscosity tends to zero. We address common practice in the
construction of numerical solutions for turbulent flows.
Applications to type Ia supernova are discussed.

Nonuniqueness (both mathematical and numerical) 
of solutions to the Euler equation is well known in the study of shock waves
and its resolution is also well known: a maximum rate of
entropy production is imposed
as a selection criteria to yield a unique and physically
relevant solution.
But nonuniqueness persists
in solutions of the incompressible Euler equation, where shock waves do not 
occur. Again, a physical principle must be added to select the physically 
meaningful solution.

This paper poses a challenge to existing standards of verification and
validation (V\&V).
We propose that if turbulence is present in the
problem solved, standards of V\&V
should ensure the physical relevance of the solutions.

As with the shock wave example, inadmissible numerical solutions of
turbulent phenomena are also possible. We identify three broad classes 
of numerical solutions to the problems of Rayleigh-Taylor (RT) turbulent mixing
and compare them to experimental data \cite{SmeYou87}. 
one of these agrees with the data, while two do not.
The second main result of this
paper is to identify these other two, solutions that include no subgrid
terms and those for which the subgrid terms are limited, i.e., the 
Implicit Large Eddy Simulation (ILES), as physically inadmissible
solutions of the turbulent RT mixing data \cite{SmeYou87}.
ILES and solutions which report a DNS status and lack subgrid terms.
These latter two solutions do not agree with each other,
further indicating nonuniqueness issues.

To account for observed discrepancies between ILES predictions and 
experimental data, it is common to add ``noise'' to the physics model.
As noise increases the entropy, some discrepancies between simulation
and measured data are removed. 

The solution with noise is, however, not predictive. Not only
can it be missing in the required amounts, but it is only a qualitative cure, 
with no defined noise level or noise frequency spectrum specified.

The maximum entropy rate is a clearly defined physics principle. We
propose it as a solution to the Euler equation nonuniqueness problem.

Reynolds averaged Navier Stokes (RANS) simulations resolve all length
scales needed to specify the problem geometry.
Large eddy simulations (LES) not only
resolve these scales, but in addition they resolve some, but not 
all, of the generic turbulent flow. The mesh scale, i.e., the finest of the
resolved scales, 
occurs within the turbulent flow. As this is a strongly coupled flow
regime, problems occur at the mesh cutoff. Resolution of all relevant
length scales, known as Direct Numerical Simulation (DNS) is 
computationally infeasible for many problems of scientific and
technological interest. As a consequence, an understanding of the
problems and opportunities of LES is an important issue.

The subgrid scale
(SGS) flow exerts an influence on the flow at the resolved level.
Because this SGS effect
is not part of the Navier-Stokes equations,
additional modeling terms are needed in the equations. These
SGS terms added to the right hand side (RHS) of the
momentum and species concentration equations
generally have the form
\begin{equation}
\label{eq:sgs}
\nabla\nu_t \nabla \quad  {\mathrm{and}} \quad \nabla D_t \nabla \ .
\end{equation}
The coefficients $\nu_t$ and $D_t$ are called eddy viscosity and eddy
diffusivity.

According to ideas of Kolmogorov \cite{Kol41}, the energy in a turbulent
flow, conserved, is passed in a cascade from larger vortices to smaller ones.
This idea leads to the scaling law \cite{Kol41}
\begin{equation}
\label{eq:K41}
\langle |v(k)|^2 \rangle = C_K \epsilon^{2/3} |k|^{-5/3}
\end{equation}
for the Fourier coefficient $v(k)$ of the velocity $v$. Here
$C_K$ is a numerical coefficient and
$\epsilon$, the energy dissipation rate,  denotes the rate at which the energy
is transferred within the cascade.
It is a measure of the intensity of the turbulence.

At the grid level, the numerically modeled 
cascade is broken. The role of the
SGS terms is to dissipate this excess grid level energy so that the
resolved scales see a diminished effect from the grid cutoff.
This analysis motivates the SGS coefficient $\nu_t$, while a conservation
law for species concentration
similarly motivates the coefficient $D_t$.

Higher order compact schemes may omit any subgrid model
in their study of RT mixing. As an example,
\cite{CabCoo06} present a nominally DNS solution,
which, however,  is not validated by comparison to experiment.
Moreover, the DNS characterization
of the simulation is not documented, with $D$ and $\nu$ not specified.
It appears from the text that DNS refers to globally defined solution
parameters such as the globally defined Kolmogorov scale $\eta$ in relation
to the mesh spacing, with $\nu$ and $D$ defined on this basis.
Such resolution misses local fluctuations in 
the turbulent intensity, which require dynamically defined SGS terms
added to the equation. As \cite{CabCoo06} is focused on applications to 
supernova Ia, additional comments are placed in our SN Ia discussion.

ILES is the computational model
in which the minimum value of $\nu_t$ is chosen so that a minimum of  grid level
excess energy is removed to retain the $|k|^{-5/3}$ scaling law, while
the prefactor $C_K\epsilon^{2/3}$ is not guaranteed. It thus depends on
limited and not full use of the subgrid terms that correspond to the
local values of the energy dissipation cascade. 
An ILES version of Miranda,  a modern higher order compact scheme, given in
\cite{MorOlsWhi17}, 
details in the construction of the ILES version of this code and analyzes
a number of scaling related properties of the RT solutions the 
algorithm generates. The subgrid terms are chosen
not proportional to the Laplacian as in (\ref{eq:sgs}),
but as higher power dissipation rates,
so that large wave numbers are more strongly suppressed. 
The SGS modeling coefficients $\nu_t$ and $D_t$ are chosen as global
constants. The basis for the choice is to regard the accumulation of
energy at the grid level as a Gibbs phenomena to be minimized 
\cite{MorOlsWhi17}. Miranda
achieves the ILES goal of an exact $-5/3$ spectral decay,
see Fig. 3 right frame in Ref. \cite{MorOlsWhi17}.

FronTier uses dynamic 
SGS models \cite{GerPioMoi91,MoiSquCab91},
and additionally uses a sharp interface model to reduce numerical
diffusion. In this method, 
SGS coefficients $\nu_t$ and $D_t$ are defined in terms
based on local flow conditions,
using turbulent scaling laws, extrapolated from an analysis
of the flow at one scale coarser, where the subgrid flow is known.

The philosophy and choices of the SGS terms are completely different among
the compact schemes, ILES and
FronTier, a fact which leads to differences in the obtained
solutions. Solution differences between FronTier and ILES were reviewed in
\cite{ZhaKamShe18}, with FronTier but not ILES showing agreement with the
data \cite{SmeYou87}. The schemes totally lacking SGS terms are even 
further from the experimental data \cite{SmeYou87}.

As shown in \cite{ZhaKamShe18},
long wave length noise in the initial conditions was eliminated as
a possible explanation of the discrepancies between ILES simulations and
experimental data for the RT instability growth
rate constant $\alpha_b$.
We also note the mixedness parameter measured in \cite{MueSch1_09}, is furthest
from experiment in \cite{CabCoo06}, is improved in the Miranda simulation
code \cite{MueSch1_09} lacking subgrid terms but with improved modeling of
experimental parameters, and further improved by the FronTier simulation
\cite{GliPloLim15}.

\section{Scaling laws compared}
\label{sec:scaling}

\begin{figure}
\begin{center}
\includegraphics[width=0.45\textwidth]{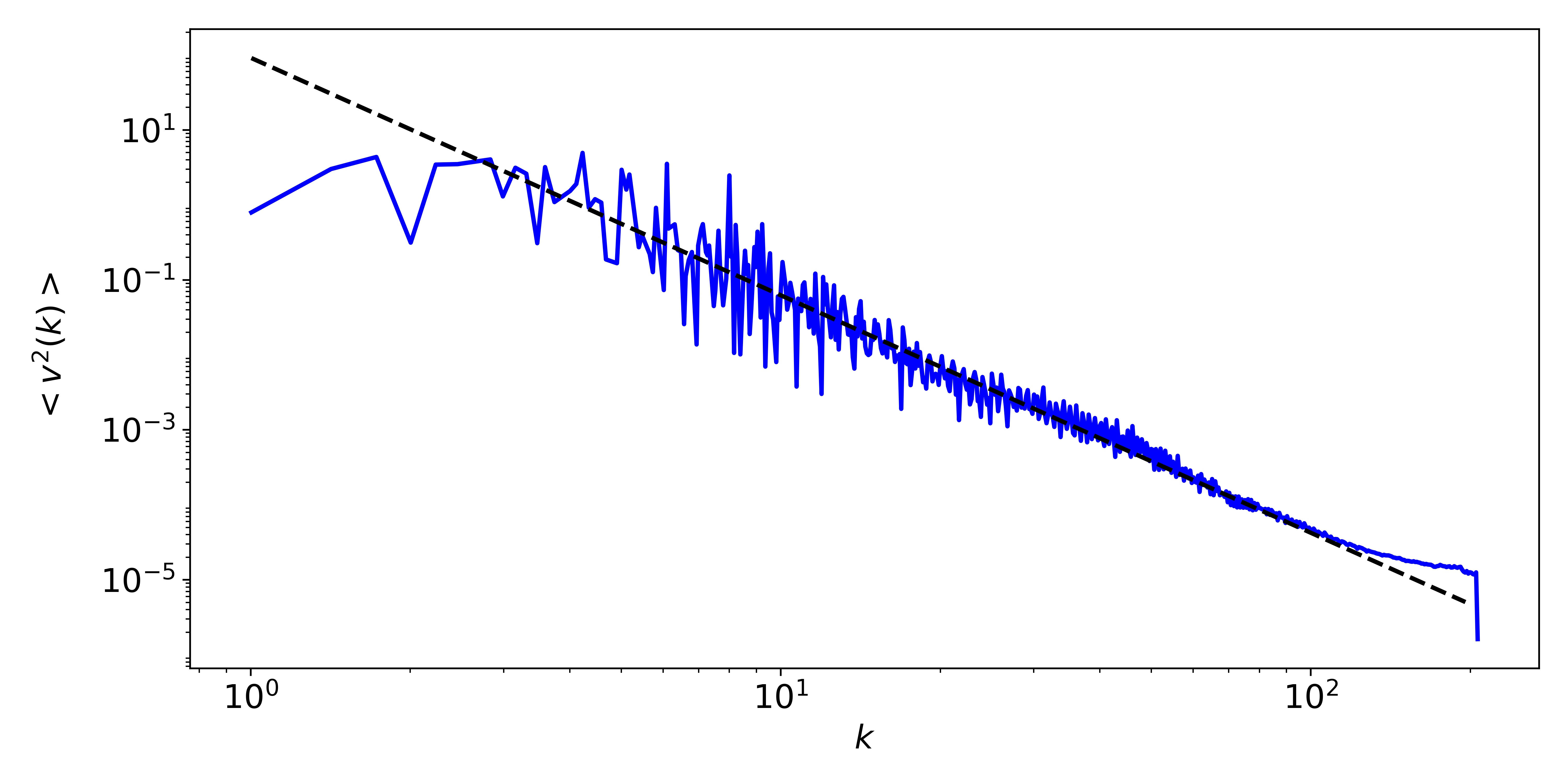}
\end{center}
\caption{
\label{fig:spectral}
Plot of the spectral decay rate, in log log variables, from the
two point function in log variables (as studied in \cite{Mah17}).
Numerical data from the final time step 
RT simulations of experiment 105 reported in
\cite{ZhaKamShe18}.
The immiscible decay rate -3.17 reflects a combination of
turbulent intermittency and the effects of a stirring cascade.
}
\end{figure}

Here we focus on differences in the spectral scaling exponents. As
\cite{CabCoo06,MorOlsWhi17} employ a thinly diffused initial layers
separating two fluids of distinct densities,
the immiscible experiments of \cite{SmeYou87} are the most appropriate for
comparison. \cite{CabCoo06} does not report velocity spectral
scaling properties,
but this reference does report the very large growth of the interfacial mixing 
area, \cite{CabCoo06} Fig. 6, 
a phenomena which we have also observed \cite{LeeJinYu07,LimYuJin07}.
The scaling rate we observe, Fig.~\ref{fig:spectral} 
from the late time 
FronTier simulations reported in \cite{ZhaKamShe18}, shows
a strong decay rate in the velocity spectrum, 
resulting from a combination of the turbulent fractal decay 
and a separate cascading process we refer to stirring.
Stirring is the mixing of distinct regions in a two phase flow. It occurs in the
concentration equation and is driven by velocity fluctuations. For stirring,
the concentration equation describes the
(tracked) front between the phases. Stirring fractal behavior is
less well studied than turbulent velocity. It 
accounts for the very steep velocity spectral decay seen in 
Fig.~\ref{fig:spectral}. In contrast, ILES \cite{MorOlsWhi17} captures
neither the expected turbulent intermittency correction to  the decay rate nor
any stirring correction beyond this.

We summarize in Table~\ref{table:compare}
the major code comparisons of this paper, 
based on the RT instability growth rate $\alpha_b$.
A compact, higher order scheme \cite{CabCoo06} has the
smallest value $\alpha_b$. ILES is larger, and the FronTier
scheme using dynamic SGS is the largest of the three,
and in agreement with experiment.
\begin{table}
\caption{
\label{table:compare}
Three types of RT simulation algorithms according to their treatment of
SGS terms and their value for $\alpha_b$, compared
to the data of \cite{SmeYou87}.
}
\begin{centering}
\begin{tabular}{|l|l|l|c|}
\hline
Code			& SGS terms 	& solution 	& evaluation  	\\
			&		& properties	& relative to \cite{SmeYou87} \\
\hline
\hline
compact high	& No SGS		& $\alpha_b \sim 0.02$	& Inconsistent\\
order \cite{CabCoo06}	&		& 			& 	\\
\hline
Miranda			& Limited SGS	& $\alpha_b \sim 0.03$	& Inconsistent\\
ILES \cite{MorOlsWhi17}	&		& 			&             \\
\hline
FronTier		& Dynamic	& $\alpha_b \sim 0.06$	& Consistent  \\
\cite{ZhaKamShe18}	& SGS		& 			&             \\
\hline
\end{tabular}
\end{centering}
\end{table}

\section{Maximum entropy production rate}
\label{sec:max-entropy}

Our first main result is to establish a plausible argument
for the validity of the maximum entropy 
production rate for Euler equation turbulence.
The admissibility condition is an extension of the second law of 
thermodynamics, in the sense that under this extension,
the physically admissible dynamic processes are constrained
more tightly than those allowed by the second law itself.
It has
 been applied successfully to many natural processes 
 \cite{MarSel06,MihFarPai17} 
including problems in climate science (terrestrial and other planets) \cite{OzaOhmLor03}, in
astrophysics, and the clustering of galaxies. As
noted in \cite{KleDyk10}, it does not have the status of an
accepted law of physics. 
A fundamental obstacle to validation of this
principle can be seen in the lack of a variational principle which combines
conservative and dissipative processes.

We avoid this 
fundamental question, and more narrowly outline a possible validation of the
maximum entropy principle
in the context of Euler equation turbulence.
The variational principle  we find, in this context,
specifies  an extreme value for the entropy production.
As this is applied at each infinitesimal increment
of time, the maximum entropy production
principle actually guarantees a maximal rate of
entropy production. For thermal processes, such a law is well validated,
and leads to the phenomenological Fourier law for thermal conductivity.

According to multifractal theories of turbulence \cite{Fri95}, turbulence 
is intermittent, with intense regions of turbulence occurring in
clusters. There is a further clustering of clusters, a process which
continues to all orders. These higher order clusters are defined
in terms of structure functions, to be introduced in Sec.~\ref{sec:vel-ent}.
Before getting into technical details, we emphasize the
central modeling assumptions that make the maximum entropy principle
valid.  For each order $p$ of clustering, a fractal set is
defined. Given a length scale $l$, the fractal set at this length scale has
a measure which is exponentially small in $l$. The central physics modeling
assumption for fractal turbulence is
\begin{itemize}
\item
(Fractal)
All the energy for the $p$ level of clustering 
is contained in a small fractal set $x_p$,
realized at the length scale $l$ as the set $x_{p,l}$.
The energy on the set, $E_{p,l}$, (defined at the scale $l$) is a constant.
\end{itemize}
This modeling assumption is used in the analysis of power laws
and Poisson processes describing the beta model of
Euler equation turbulence \cite{Fri95}.
It follows that the steady state energy dissipation and entropy production
of the order $p$ clustering
at length scale $l$ are given by
\begin{equation}
\label{eq:entropy-def}
E_{p,l} \int_{x_{p,l}} x dx \quad {\mathrm{and}} \quad
E_{p,l} \int_{x_{p,l}} x \ln x dx \ ,
\end{equation}
We observe that the energy occurs 
outside of the integrals,
and that the term $(1-x) \ln (1-x)$ is missing from the entropy.
To model a time
dependent state which has not yet achieved equilibrium, and is still
evolving in time, the only change to
(\ref{eq:entropy-def}) is that the equations are
multiplied by the fractional equilibrium part of the state.

The log Poisson model \cite{SheLev94} selects a fractal set
to describe each order $p$ of clustering. The choice, conditional on
prior choices for smaller $p$ values, is not defined by an exponential, i.e., a
pure fractal, but a mixture of exponentials in the energy dissipation rates.
As the mixture is not
narrowly concentrated about its peak value, the applicability of
hypothesis 
(Fractal) cannot be assumed. In the limit of large $p$, however,
the mixture of exponentials is narrow, so that (Fractal) is justified
for physically realizable solutions of Euler equation turbulence.
The peak values for finite $p$ are not identified in the log Poisson
analysis, which finds the mean of the mixture exponentials on the
basis a universality hypothesis. This multifractal model,
evaluated for large $p$ is
applied uniformly to all $p$. From the excellent agreement of these predictions
with multiple experiments and simulations (1\% accuracy) \cite{SheLev94},
the log Poisson model is validated. 
A plausible principle to select the physically relevant solutions from among
the multiple nonunique solutions of the Euler equation, 
suggested by this analysis, is the principle of
maximal rate of energy dissipation. The analysis of \cite{SheLev94} maximizes
the mean value of competing exponentials rather than their peak value. 
The mean and peak coincide in the limit of large $p$, but the distinction
between them for finite $p$ is a gap remaining in any validation argument.
The maximum energy dissipation rate is a viable candidate for the
required selection principle among nonunique solutions of the Euler equation.
Accepting this, our analysis will be complete with solutions lacking
subgrid terms and ILES seen to be invalid physically. 
To the extent that some maximal entropy likelihood reasoning is applicable,
for example such as (Fractal), a  maximum entropy production rate principle
for the selection of physically relevant solutions of the Euler equation
for fully developed turbulence would follow.

We refer to the highly
developed extensions \cite{StJ05,DubGra96,DubGra96a,SheWay95}
of \cite{SheLev94}. The references \cite{DubGra96,DubGra96a,SheWay95}
extend the log Poisson model to continuous $p > 0$. These references
do not resolve the issue
of either a maximum entropy production rate or a maximum energy dissipation
rate for fully developed turbulence, but they appear to offer a plausible
route for possible validation of either of these.

The dynamic equations are of Fokker-Plank type. 
The dissipation operator is a sum of a conventional Laplacian, for the
thermal diffusion and an integral over $p > 0$ of the order $p$ clustering
contribution, which is a fractal, or power law dissipation, expressed in
powers of the length scale $l$.

\section{The second main result}
\label{sec:results}
In view of these observations, we note three
independent reasons for concluding that the absence of subgrid
terms or their limited presence in ILES is problematic 
on physical grounds. 
\begin{enumerate}
\item The two limited subgrid schemes do not satisfy
the maximum entropy production rate principle.
\item The two limited subgrid schemes are
in violation of incontrovertible experimental and
simulation evidence that the true total spectral 
decay rate is more negative than $-5/3$ \cite{Fri95}.
\item These two schemes  understate the dissipated energy 
and are thus unphysical.
\end{enumerate}
These are logically independent statements. The order is decreasing in the
fundamental nature of the statement and increasing in the simplicity of the
assumptions.  Any one of these points is  sufficient to invalidate
schemes lacking in subgrid terms or ILES , with limited subgrid terms.
Point 1 is the most fundamental in nature, and it is the
subject of the remainder of this paper. Point 2
rests on established laws of physics and assumes the relevance of
Kolmogorov scaling laws with their intermittency corrections to RT mixing.
Point 3 assumes nothing. Simulations, even ILES simulations, show a transfer
of energy from large to small scales. Point 3 accepts this as a physical
fact. The energy transfer
is not a numerical feature to be minimized, but a property of the
solutions to be modeled correctly. The grid level cutoff terminates this
transfer, and point 3 notes that the transfer, from the grid level to the
subgrid level is incorrectly modeled in both types of limited dissipation
schemes.

For the reader satisfied with
either points 2 or 3, the remainder of the paper can be ignored, and the
discussion has been completed,
independent of the remainder of the paper.

The problems with current computational paradigms are well summarized by 
Zhou \cite{Zho17a}, Sec. 6
regarding evaluation of the RT
instability growth rate
$\alpha_b$, ``agreement between simulations and experiment are worse today
than it was several decades ago because of the availability of more
powerful computers.''
As our computational method depends on front tracking in addition to 
dynamic subgrid modes (which address items 1-3 above), we additionally
quote from Zhou \cite{Zho17a}, Sec. 5.2, in discussing \cite{GeoGliLi05}:
``it was clear that accurate numerical tracking to control numerical mass 
diffusion and accurate modeling of physical scale-breaking phenomena
surface tension were the critical steps for the simulations to agree with
the experiments of Read and Smeeton and Youngs''.

We raise the possibility of ILES related errors in an
analysis of the deflagration to detonation transition in type
Ia supernova. In that these simulations depend on ILES,
their predictive value may be questioned.
We propose a simple simulation search method for
rare events, in which a physics simulation code drives the turbulence 
modeling. In agreement with \cite{CabCoo06}, we recommend a new 
class of turbulent combustion subgrid models. See Sec.~\ref{sec:Ia}.

\subsection{Rayleigh-Taylor turbulent mixing}
\label{sec:RT}

We assess ILES in terms of the RT instability of 
acceleration driven instabilities, and the prediction of the growth
rate $\alpha_b$ of this instability.  
We identify situations in which ILES is in near agreement with
experimental measurement of the growth rate
in this measure \cite{Mue08,MueSch1_09}, and ones 
\cite{DimYou04} where its predictions  differ
by a factor of about 2 from experiments \cite{SmeYou87}.
The first case is characterized by 
\begin{itemize}
\item{(a)} low levels of turbulence 
\item{(b)} high levels of long wave length perturbations 
(``noise'') in the initial
conditions and \item{(c)} diffusive parameters in the physics model.
\end{itemize}
Regarding item (c), we observe that the successful ILES simulations
referenced above concerned hot-cold water, with a moderate Schmidt
number of 7, whereas no results are reported for the very low diffusive
fresh-salt water channel with a Schmidt number of 600.

\subsection{Noise as an adjustable parameter}

The postulate \cite{You03} of noise in the initial data
\cite{SmeYou87}
was shown to lead to agreement of the ILES predictions with experiment.
In previous studies \cite{GliShaKam11,ZhaKamShe18}, we have shown that this
postulate is not valid.
The long wave length noise is present,
but with a sufficiently small amplitude that its influence on the
instability growth rate is about $5\%$. Thus long wave length
initial ``noise'' in the initial conditions for \cite{SmeYou87}
is not sufficient to account
for the factor of 2 discrepancy between ILES and this data. 

We regard ``noise'' as a palliative, and not a fundamental principle.
The noise level is not specified, nor is its frequency spectrum,
so that standards of predictive
science are not met. As noted, ``noise'', of the required intensity,
is missing in some instances. We propose the maximum entropy production
rate as a more satisfactory solution to the problem of Euler equation
turbulence nonuniqueness.

ILES simulations have been used in the study of incompressible turbulence,
a problem with ample experimental data reviewed in \cite{SheLev94}.
In such simulations,
``noise'' is added to the initial conditions. In this case the high
frequency component of the noise is important. Agreement with experiment
is obtained. Pure ILES, with no added noise would not meet this test.

\subsection{Outline of derivation}
\label{sec:outline}

Our reasoning is based on three fundamental laws of physics:
\begin{itemize}
\item Conservation of energy, the first law of thermodynamics
\item Maximum entropy production rate, an extension of the second law of
thermodynamics
\item Universality in the clustering and compound clustering of intermittency
in fully developed turbulence.
\end{itemize}
The third item is formulated in \cite{SheLev94}. 
In the multifractal description of turbulence, universality
states that the compound clusters, that is the multiple fractals
in the description of turbulent intermittency, must all obey a common
law. There can be no new physical law or parameter in passing
from one level of clustering to the next. The law is evaluated
in closed form \cite{SheLev94} in the limit that the order of the clustering
becomes infinite.
It is a power scaling law. By universality, this
law is then applied to clustering at all orders.

The universality theories are developed for single constant density
incompressible turbulence. Our use in a variable
density context is an extrapolation of these theories
beyond their domain of strict validated applicability. Scaling
laws are similarly extrapolated. Such extrapolations are widely
used (and verified) in simulation studies. For convenience, 
the Reynolds stress analysis uses this approximation.

In shock wave modeling, the Euler equation shock
wave introduces a Gibbs phenomena of overshoot. The instability resulting
is removed by dissipation (artificial viscosity, and its modern
variants) of the minimum amount to just prevent the overshoot.
The turbulent cascade of energy is not a Gibbs phenomena. It is
an observable fact and not a numerical artifact.
Minimizing its magnitude is an error, as opposed to
an accurate model of the mesh dissipated error in the dynamic SGS models.

We proceed in the following steps. Using the Reynolds stress, we 
express the SGS terms to be modeled as a truncated two point function.
In this formulation, we identify the minimum (ILES) and maximum
(dynamic SGS) alternatives. 

We then proceed
from velocity fluctuations 
to the energy dissipation rate $\epsilon$ and from the latter to
the entropy production rate. At each step we are looking at truncated
two point functions. At the end, we are looking at the
entropy production rate and must choose the 
solution with maximum entropy production rate.

Each step is monotone and preserves the minimum-maximum choice.
Reasoning backwards, we see that the maximum
choice is needed at the outset, and so ILES is inadmissible.

The transition, from velocities to energy truncated two point functions,
has two components. The first is a scaling analysis to show equivalence,
but in the process the order of clustering changes. The second component
in this transition
is to apply universality: all orders of clustering must obey a common
minimum-maximum choice. 
.

\subsection{From velocities to entropy}
\label{sec:vel-ent}

\subsubsection{Reynolds stress}
The Reynolds stress results from regarding the mesh values as
cell averaged quantities. This creates an obvious problem for nonlinear
terms of the Euler equation. From the momentum equation, the quadratic
nonlinearity is replaced by the product of the cell mean values. The
resulting error, transferred to the RHS of the momentum equation is the
negative of the gradient of the Reynolds stress, defined as
\begin{equation}
\label{eq:rey}
R = \overline{ v^2} - \overline{v} ~ \overline{v}
\end{equation}
in the case of constant density, with a more complex expression involving
density weighted (Favre) averages in the variable density case.

The added force term $-\nabla R$ on the right hand side (RHS)
of the momentum equation is modeled
as $\nu_t \Delta v$. Thus we see that the minimum and maximum values for
the energy dissipation rate $\nu_t$ correspond to
minimum and maximum values for models of $-\nabla R$. $R$, as a truncated
two point function, vanishes as its argument becomes infinite and is peaked
at the origin. Thus minimum and maximum values for $-\nabla R$ 
correspond to minimum and maximum values for $R$ itself.

\subsubsection{Velocities to energy}

As technical preparation for the analysis of this section, we define the 
structure functions. They make precise the intuitive picture 
of multiple orders of clustering for intermittency.
There are two families of structure functions, one for 
velocity fluctuations  and the other for  the energy
dissipation rate $\epsilon$. The structure functions are the expectation
value of the $p^{th}$ power of the variable.  For each value of $p$,
they define a fractal and satisfy a power law in their decay in a scaling 
variable $l$. The structure functions and the associated scaling exponents
$\zeta_l$ and $\tau_l$ are defined as
\begin{equation}
\label{eq:zeta-tau}
\langle \delta v_l^p \rangle \sim l^{\zeta_p}
\quad {\mathrm{and}} \quad
\langle \epsilon_l^p \rangle  \sim l^{\tau_p}
\end{equation}
where $\delta v_l$ and $\epsilon_l$ are respectively
the averages of velocity differences and
of $\epsilon$ over a ball of size $l$.
The two families of exponents are related by a simple scaling law
\begin{equation}
\label{eq:zeta-tau2}
\zeta_p = p/3 + \tau_{p/3}
\end{equation}
derived on the basis of scaling laws and dimensional analysis \cite{Kol62}.
This would seem to accomplish the velocity fluctuation 
to energy dissipation rate
step, preserving the minimum vs. maximum choice,
but it does not, because 
the value of $p$ to which it applies has changed.

To fill this gap, we turn to the assumption of universality formulated
in terms of the $\tau_p$ \cite{SheLev94}, and as explained
with mathematical formalisms replacing some of the reasoning of a  theoretical
modeling nature, \cite{SheWay95,DubGra96,DubGra96a,SheZha09,LiuShe03}.
As a function of $p$,
$\tau_p$ is a fractional order cubic, defined in terms of a fractional
order dissipative operator with a fractional order exponent $\beta$.
This relation is derived exactly in the limit as $p \rightarrow \infty$,
and in the name of universality, then applied to all values of $p$.

As a monotone fractional order cubic, it follows that the minimum-maximum choice
for any $p$ is reflected in the same choice for all $p$. We have thereby
completed the velocity to energy dissipation rate step, and
preserved the minimum vs. maximum choice.

From the modeling principle (Fractal), the
energy dissipation rate is maximized exactly when the
entropy production rate is maximized.
The maximum choice for the entropy production rate is required
and the minimum choice is inadmissible. Reasoning backwards to the
original energy dissipation choices, the minimum rate of energy
dissipation (ILES) is inadmissible.

\section{Significance: an example}
\label{sec:Ia}

For simulation modeling of turbulent flow nonlinearly coupled to other
physics (combustion and reactive flows, particles embedded in turbulent flow,
radiation), the method of dynamic SGS turbulent flow models, which only
deals with average subgrid effects, may be insufficient. In such cases,
the turbulent fluctuations or the full two point correlation function
is a helpful component of SGS modeling. Such a goal is only partially
realized in the simplest of cases, single density incompressible 
turbulence. For highly complex physical processes, the knowledge of the
domain scientist must still be retained, and it appears to be
more feasible to bring
multifractal modeling ideas into the domain science communities.

In this spirit, we propose here a simple method
for the identification of (turbulence related) extreme events
through a modification of adaptive mesh refinement (AMR), which we 
call Fractal Mesh Refinement (FMR). We propose FMR to seek a
deflagration to detonation transition (DDT) 
in type Ia supernova. 

FMR  allows high levels of strongly focused resolution.
The method is proposed to assess the extreme events generated
by multifractal turbulent nuclear deflagration. Such events, in a white
dwarf type Ia supernova progenitor, are assumed to lead to DDT,
which produces the observed type Ia supernova.
See \cite{ZinAlmBar17,CalKruJac12} and references cited there.

FMR refines the mesh not adaptively where needed,
but only in the most highly critical regions where most important, and
thereby may detect
DDT trigger events within large volumes at a feasible computational cost.

The detailed mechanism for DDT is presumed to be diffused
radiative energy
arising from some local combustion event of extreme
intensity, in the form of a convoluted flame front, embedded in
a nearby volume of unburnt stellar material close to
ignition.
Consistent with the Zeldovich theory \cite{Lee08}, a wide spread
ignition and explosion may result. 
FMR refinement criteria will search for such
events. In this plan, the FMR search should avoid ILES.
See \cite{Gli18}.

There is a minimum length scale for wrinkling of a turbulent
combustion front, called the Gibson scale. Mixing can proceed in the
absence of turbulence, via stirring. Thus the Gibson scale is not the
correct limiting scale for a DDT event. Stirring, for a flame front,
terminates at a smaller scale, the width of the flame itself. The analysis of
length scales must also include correctly modeled transport for charged
ions \cite{MelLimRan14},
which can be orders of magnitude larger than those inferred
from hydro considerations. The 
microstructure of mixing for a flame front could be thin flame regions
surrounded by larger regions of burned and unburned stellar material
(as with a foam of soap bubbles, with a soap film between the bubbles).
Here again multifractal and entropy issues appear to be relevant,
although not subject to theoretical analysis comparable to that
of multifractal for turbulent flow.
A multifractal clustering of smaller bubbles separated by flame fronts
can be anticipated, and
where a sufficient fraction of these bubbles are unburnt stellar material,
a trigger for DDT could occur.

FMR, with its narrow focus on extreme events, will come closer to discovering
such DDT triggers than will an AMR algorithm design.
For this purpose, the astrophysics code should be based on dynamic subgrid
SGS, not on ILES. 

We return to the discussion of \cite{CabCoo06}. Our FronTier computations
of a 2D interface surface length  are in qualitative agreement with those of
\cite{CabCoo06} for the surface area.
Such models of interface area should be the basis for subgrid scale modeling
of the turbulent flame intensity. Work is currently in progress to
construct an experimentally validated subgrid scale microstructure to 
complement models of turbulent flame surface area.  These subgrid
models may play a role in reaching beyond length scales reachable by FMR.

\section{Conclusions }
\label{sec:conc}

We have shown that the ILES algorithm for the solution of Euler equation 
turbulence is inadmissible physically. It is in violation of the
physical principle of maximum rate of entropy production.

We have explained observations of experimental flows for which this 
error in ILES has only a minor effect. They are associated with high levels
of noise in the initial conditions, low levels of turbulent intensity
and diffusive flow parameters.
Prior work, e.g., \cite{GliShaKam11,GeoGliLi05,ZhaKamShe18}
pertain to simulation validation studies RT instability
experiments with a stronger intensity of turbulence
and for which such significant long wave length perturbations to the
initial data are missing. In these experiments, the present analysis 
provides a partial explanation for the factor of about 2 discrepancy between
observed and ILES predicted instability growth rates. 

We have noted the potential for ILES related errors to influence
ongoing scientific investigations, including the search for
DDT in type Ia supernova.

We believe V\&V standards should include an analysis of the physical
relevance of proposed solutions to flow problems, specifically turbulent
and stirring problems.
The ILES simulations of the experiments of \cite{SmeYou87} fail this
test by a factor of 2 in the RT growth rate $\alpha_b$, and
on this basis we judge them to be physically inadmissible. 

We recognize that the conclusions of this paper will be controversial within 
the ILES and high order compact turbulent simulation communities. 
A deeper consideration of the
issues raised here is a possible outcome. 
The issues to be analyzed are clear:
\begin{itemize}
\item
Is the transport of energy and concentration, blocked at the
grid level, to be ignored entirely \cite{CabCoo06}? 
\item
Is it to be regarded as a Gibbs phenomena \cite{MorOlsWhi17},
and thus to be minimized? 
\item
Is it a physical phenomena,
to be modeled accurately \cite{GerPioMoi91,MoiSquCab91}?
\end{itemize}
If the response to this paper is an appeal to 
consensus (everyone else is doing it),
the argument fails. Consensus is of course a weak argument, and
one that flies  in the face of standards of V\&V. More significantly,
there is a far larger engineering community
using dynamic SGS models in the design of engineering structures
tested in actual practice.
This choice is backed by nearly three decades of
extensive experimental validation.  It is further used
to extend the calibration range of RANS simulations beyond available
experimental data. The resulting RANS, calibrated to dynamic SGS LES data,
are widely used in the design and optimization
of engineering structures; these are also tested in real applications.
Consensus in this larger community overwhelms the ILES consensus
by its shear magnitude, and ILES loses the consensus argument.

\section{Acknowledgements}
\label{ack}

Use of computational support by the Swiss National Supercomputing Centre is gratefully acknowledged. 
Los Alamos National Laboratory Preprint LA-UR-18-30837.


\begin{thebibliography}{38}
\expandafter\ifx\csname natexlab\endcsname\relax\def\natexlab#1{#1}\fi
\expandafter\ifx\csname bibnamefont\endcsname\relax
  \def\bibnamefont#1{#1}\fi
\expandafter\ifx\csname bibfnamefont\endcsname\relax
  \def\bibfnamefont#1{#1}\fi
\expandafter\ifx\csname citenamefont\endcsname\relax
  \def\citenamefont#1{#1}\fi
\expandafter\ifx\csname url\endcsname\relax
  \def\url#1{\texttt{#1}}\fi
\expandafter\ifx\csname urlprefix\endcsname\relax\def\urlprefix{URL }\fi
\providecommand{\bibinfo}[2]{#2}
\providecommand{\eprint}[2][]{\url{#2}}

\bibitem[{\citenamefont{{De Leliss} and Szekelyhidi}(2009)}]{DelSze09}
\bibinfo{author}{\bibfnamefont{C.}~\bibnamefont{{De Leliss}}} \bibnamefont{and}
  \bibinfo{author}{\bibfnamefont{L.}~\bibnamefont{Szekelyhidi}},
  \bibinfo{journal}{Ann. Math.} \textbf{\bibinfo{volume}{170}},
  \bibinfo{pages}{1471} (\bibinfo{year}{2009}).

\bibitem[{\citenamefont{{De Leliss} and Szekelyhidi}(2010)}]{DelSze10}
\bibinfo{author}{\bibfnamefont{C.}~\bibnamefont{{De Leliss}}} \bibnamefont{and}
  \bibinfo{author}{\bibfnamefont{L.}~\bibnamefont{Szekelyhidi}},
  \bibinfo{journal}{Arch. Rat. Mech. Anal.} \textbf{\bibinfo{volume}{195}},
  \bibinfo{pages}{225} (\bibinfo{year}{2010}).

\bibitem[{\citenamefont{Smeeton and Youngs}(1987)}]{SmeYou87}
\bibinfo{author}{\bibfnamefont{V.~S.} \bibnamefont{Smeeton}} \bibnamefont{and}
  \bibinfo{author}{\bibfnamefont{D.~L.} \bibnamefont{Youngs}},
  \bibinfo{type}{AWE Report Number} \bibinfo{number}{0 35/87}
  (\bibinfo{year}{1987}).

\bibitem[{\citenamefont{Kolmogorov}(1941)}]{Kol41}
\bibinfo{author}{\bibfnamefont{A.~N.} \bibnamefont{Kolmogorov}},
  \bibinfo{journal}{Doklady Akad. Nauk. SSSR} \textbf{\bibinfo{volume}{30}},
  \bibinfo{pages}{299} (\bibinfo{year}{1941}).

\bibitem[{\citenamefont{Cabot and Cook}(2006)}]{CabCoo06}
\bibinfo{author}{\bibfnamefont{W.}~\bibnamefont{Cabot}} \bibnamefont{and}
  \bibinfo{author}{\bibfnamefont{A.}~\bibnamefont{Cook}},
  \bibinfo{journal}{Nature Physics} \textbf{\bibinfo{volume}{2}},
  \bibinfo{pages}{562} (\bibinfo{year}{2006}).

\bibitem[{\citenamefont{Morgan et~al.}(2017)\citenamefont{Morgan, Olson, White,
  and McFarland}}]{MorOlsWhi17}
\bibinfo{author}{\bibfnamefont{B.~E.} \bibnamefont{Morgan}},
  \bibinfo{author}{\bibfnamefont{B.~J.} \bibnamefont{Olson}},
  \bibinfo{author}{\bibfnamefont{J.~E.} \bibnamefont{White}}, \bibnamefont{and}
  \bibinfo{author}{\bibfnamefont{J.~A.} \bibnamefont{McFarland}},
  \bibinfo{journal}{J. Turbulence} \textbf{\bibinfo{volume}{18}}
  (\bibinfo{year}{2017}).

\bibitem[{\citenamefont{Germano et~al.}(1991)\citenamefont{Germano, Piomelli,
  Moin, and Cabot}}]{GerPioMoi91}
\bibinfo{author}{\bibfnamefont{M.}~\bibnamefont{Germano}},
  \bibinfo{author}{\bibfnamefont{U.}~\bibnamefont{Piomelli}},
  \bibinfo{author}{\bibfnamefont{P.}~\bibnamefont{Moin}}, \bibnamefont{and}
  \bibinfo{author}{\bibfnamefont{W.~H.} \bibnamefont{Cabot}},
  \bibinfo{journal}{Phys. Fluids A} \textbf{\bibinfo{volume}{3}},
  \bibinfo{pages}{1760} (\bibinfo{year}{1991}).

\bibitem[{\citenamefont{Moin et~al.}(1991)\citenamefont{Moin, Squires, Cabot,
  and Lee}}]{MoiSquCab91}
\bibinfo{author}{\bibfnamefont{P.}~\bibnamefont{Moin}},
  \bibinfo{author}{\bibfnamefont{K.}~\bibnamefont{Squires}},
  \bibinfo{author}{\bibfnamefont{W.}~\bibnamefont{Cabot}}, \bibnamefont{and}
  \bibinfo{author}{\bibfnamefont{S.}~\bibnamefont{Lee}},
  \bibinfo{journal}{Phys. Fluids A} \textbf{\bibinfo{volume}{3}},
  \bibinfo{pages}{2746} (\bibinfo{year}{1991}).

\bibitem[{\citenamefont{Zhang et~al.}(2018)\citenamefont{Zhang, Kaman, She,
  Cheng, Glimm, and Sharp}}]{ZhaKamShe18}
\bibinfo{author}{\bibfnamefont{H.}~\bibnamefont{Zhang}},
  \bibinfo{author}{\bibfnamefont{T.}~\bibnamefont{Kaman}},
  \bibinfo{author}{\bibfnamefont{D.}~\bibnamefont{She}},
  \bibinfo{author}{\bibfnamefont{B.}~\bibnamefont{Cheng}},
  \bibinfo{author}{\bibfnamefont{J.}~\bibnamefont{Glimm}}, \bibnamefont{and}
  \bibinfo{author}{\bibfnamefont{D.~H.} \bibnamefont{Sharp}},
  \bibinfo{journal}{Pure and Applied Mathematics Quarterly}
  (\bibinfo{year}{2018}), \bibinfo{note}{in press; Los Alamos National
  Laboratory preprint LA-UR-18-22134}.

\bibitem[{\citenamefont{Mueschke and Schilling}(2009)}]{MueSch1_09}
\bibinfo{author}{\bibfnamefont{N.}~\bibnamefont{Mueschke}} \bibnamefont{and}
  \bibinfo{author}{\bibfnamefont{O.}~\bibnamefont{Schilling}},
  \bibinfo{journal}{Physics of Fluids} \textbf{\bibinfo{volume}{21}},
  \bibinfo{pages}{014106 1} (\bibinfo{year}{2009}).

\bibitem[{\citenamefont{Glimm et~al.}(2015)\citenamefont{Glimm, Plohr, Lim, Hu,
  and Sharp}}]{GliPloLim15}
\bibinfo{author}{\bibfnamefont{J.}~\bibnamefont{Glimm}},
  \bibinfo{author}{\bibfnamefont{B.}~\bibnamefont{Plohr}},
  \bibinfo{author}{\bibfnamefont{H.}~\bibnamefont{Lim}},
  \bibinfo{author}{\bibfnamefont{W.}~\bibnamefont{Hu}}, \bibnamefont{and}
  \bibinfo{author}{\bibfnamefont{D.~H.} \bibnamefont{Sharp}},
  \bibinfo{journal}{Annals of Mathematical Sciences and Applications}
  (\bibinfo{year}{2015}), \bibinfo{note}{{S}tony {B}rook {U}niversity Preprint
  Number SUNYSB-AMS-15-05 and Los Alamos National Laboratory LAUR Number
  LA-UR-12-26149}.

\bibitem[{\citenamefont{Lee et~al.}(2008)\citenamefont{Lee, Jin, Yu, and
  Glimm}}]{LeeJinYu07}
\bibinfo{author}{\bibfnamefont{H.}~\bibnamefont{Lee}},
  \bibinfo{author}{\bibfnamefont{H.}~\bibnamefont{Jin}},
  \bibinfo{author}{\bibfnamefont{Y.}~\bibnamefont{Yu}}, \bibnamefont{and}
  \bibinfo{author}{\bibfnamefont{J.}~\bibnamefont{Glimm}},
  \bibinfo{journal}{Phys. Fluids} \textbf{\bibinfo{volume}{20}},
  \bibinfo{pages}{1} (\bibinfo{year}{2008}), \bibinfo{note}{{S}tony {B}rook
  {U}niversity Preprint SUNYSB-AMS-07-03}.

\bibitem[{\citenamefont{Lim et~al.}(2008)\citenamefont{Lim, Yu, Jin, Kim, Lee,
  Glimm, Li, and Sharp}}]{LimYuJin07}
\bibinfo{author}{\bibfnamefont{H.}~\bibnamefont{Lim}},
  \bibinfo{author}{\bibfnamefont{Y.}~\bibnamefont{Yu}},
  \bibinfo{author}{\bibfnamefont{H.}~\bibnamefont{Jin}},
  \bibinfo{author}{\bibfnamefont{D.}~\bibnamefont{Kim}},
  \bibinfo{author}{\bibfnamefont{H.}~\bibnamefont{Lee}},
  \bibinfo{author}{\bibfnamefont{J.}~\bibnamefont{Glimm}},
  \bibinfo{author}{\bibfnamefont{X.-L.} \bibnamefont{Li}}, \bibnamefont{and}
  \bibinfo{author}{\bibfnamefont{D.~H.} \bibnamefont{Sharp}},
  \bibinfo{journal}{Compu. Methods Appl. Mech. Engrg.}
  \textbf{\bibinfo{volume}{197}}, \bibinfo{pages}{3435} (\bibinfo{year}{2008}),
  \bibinfo{note}{{S}tony Brook University Preprint SUNYSB-AMS-07-05}.

\bibitem[{\citenamefont{Mahadeo}(2017)}]{Mah17}
\bibinfo{author}{\bibfnamefont{V.}~\bibnamefont{Mahadeo}}, \bibinfo{type}{Ph.d.
  thesis}, \bibinfo{school}{Stony Brook University} (\bibinfo{year}{2017}).

\bibitem[{\citenamefont{Martyushev and Seleznev}(2006)}]{MarSel06}
\bibinfo{author}{\bibfnamefont{L.~M.} \bibnamefont{Martyushev}}
  \bibnamefont{and} \bibinfo{author}{\bibfnamefont{V.~D.}
  \bibnamefont{Seleznev}}, \bibinfo{journal}{Phy. Reports}
  \textbf{\bibinfo{volume}{426}}, \bibinfo{pages}{1} (\bibinfo{year}{2006}).

\bibitem[{\citenamefont{Mihelich et~al.}(2017)\citenamefont{Mihelich, Faranda,
  Pailard, and Dubrulle}}]{MihFarPai17}
\bibinfo{author}{\bibfnamefont{M.}~\bibnamefont{Mihelich}},
  \bibinfo{author}{\bibfnamefont{D.}~\bibnamefont{Faranda}},
  \bibinfo{author}{\bibfnamefont{D.}~\bibnamefont{Pailard}}, \bibnamefont{and}
  \bibinfo{author}{\bibfnamefont{B.}~\bibnamefont{Dubrulle}},
  \bibinfo{journal}{Entropy} \textbf{\bibinfo{volume}{19}}
  (\bibinfo{year}{2017}).

\bibitem[{\citenamefont{Ozawa et~al.}(2003)\citenamefont{Ozawa, Ohmura,
  Lorentz, and Pujol}}]{OzaOhmLor03}
\bibinfo{author}{\bibfnamefont{H.}~\bibnamefont{Ozawa}},
  \bibinfo{author}{\bibfnamefont{A.}~\bibnamefont{Ohmura}},
  \bibinfo{author}{\bibfnamefont{R.}~\bibnamefont{Lorentz}}, \bibnamefont{and}
  \bibinfo{author}{\bibfnamefont{T.}~\bibnamefont{Pujol}},
  \bibinfo{journal}{Reviews of Geophysics} \textbf{\bibinfo{volume}{41}}
  (\bibinfo{year}{2003}).

\bibitem[{Kle(2010)}]{KleDyk10}
in \emph{\bibinfo{booktitle}{What is Maximum Entrpy Productionn and how should
  we apply it}}, edited by
  \bibinfo{editor}{\bibfnamefont{A.}~\bibnamefont{Kleidon}} \bibnamefont{and}
  \bibinfo{editor}{\bibfnamefont{J.}~\bibnamefont{Dyke}}
  (\bibinfo{publisher}{Entropy}, \bibinfo{year}{2010}), \bibinfo{note}{special
  issue, Vol. 12}.

\bibitem[{\citenamefont{Frisch}(1996)}]{Fri95}
\bibinfo{author}{\bibfnamefont{U.}~\bibnamefont{Frisch}},
  \emph{\bibinfo{title}{Turbulence: The Legacy of {A}. {N}. {K}olmogorov}}
  (\bibinfo{publisher}{Cambridge Univeristy Press},
  \bibinfo{address}{Cambridge}, \bibinfo{year}{1996}).

\bibitem[{\citenamefont{She and Leveque}(1994)}]{SheLev94}
\bibinfo{author}{\bibfnamefont{Z.~S.} \bibnamefont{She}} \bibnamefont{and}
  \bibinfo{author}{\bibfnamefont{E.}~\bibnamefont{Leveque}},
  \bibinfo{journal}{Phys. Rev. Lett.} \textbf{\bibinfo{volume}{72}},
  \bibinfo{pages}{336} (\bibinfo{year}{1994}).

\bibitem[{\citenamefont{St-Jean}(2005)}]{StJ05}
\bibinfo{author}{\bibfnamefont{P.}~\bibnamefont{St-Jean}},
  \bibinfo{journal}{Eur. Phys. J. B} \textbf{\bibinfo{volume}{46}}
  (\bibinfo{year}{2005}).

\bibitem[{\citenamefont{Dubrulle and Graner}(1996)}]{DubGra96}
\bibinfo{author}{\bibfnamefont{B.}~\bibnamefont{Dubrulle}} \bibnamefont{and}
  \bibinfo{author}{\bibfnamefont{F.}~\bibnamefont{Graner}},
  \textbf{\bibinfo{volume}{6}}, \bibinfo{pages}{817} (\bibinfo{year}{1996}).

\bibitem[{\citenamefont{Dubrulle and Granier}(1996)}]{DubGra96a}
\bibinfo{author}{\bibfnamefont{B.}~\bibnamefont{Dubrulle}} \bibnamefont{and}
  \bibinfo{author}{\bibfnamefont{F.}~\bibnamefont{Granier}},
  \textbf{\bibinfo{volume}{6}}, \bibinfo{pages}{797} (\bibinfo{year}{1996}).

\bibitem[{\citenamefont{She and Weymire}(1995)}]{SheWay95}
\bibinfo{author}{\bibfnamefont{Z.~S.} \bibnamefont{She}} \bibnamefont{and}
  \bibinfo{author}{\bibfnamefont{E.}~\bibnamefont{Weymire}},
  \bibinfo{journal}{Phys. Rev. Lett.} \textbf{\bibinfo{volume}{74}},
  \bibinfo{pages}{262} (\bibinfo{year}{1995}).

\bibitem[{\citenamefont{Zhou}(2017)}]{Zho17a}
\bibinfo{author}{\bibfnamefont{Y.}~\bibnamefont{Zhou}},
  \bibinfo{journal}{Physics Reports} \textbf{\bibinfo{volume}{720--722}},
  \bibinfo{pages}{1 } (\bibinfo{year}{2017}),
  \bibinfo{note}{http://dx.doi.org/10.1016/j.physrep.2017.07.005}.

\bibitem[{\citenamefont{George et~al.}(2006)\citenamefont{George, Glimm, Li,
  Li, and Liu}}]{GeoGliLi05}
\bibinfo{author}{\bibfnamefont{E.}~\bibnamefont{George}},
  \bibinfo{author}{\bibfnamefont{J.}~\bibnamefont{Glimm}},
  \bibinfo{author}{\bibfnamefont{X.-L.} \bibnamefont{Li}},
  \bibinfo{author}{\bibfnamefont{Y.-H.} \bibnamefont{Li}}, \bibnamefont{and}
  \bibinfo{author}{\bibfnamefont{X.-F.} \bibnamefont{Liu}},
  \bibinfo{journal}{Phys. Rev. E} \textbf{\bibinfo{volume}{73}},
  \bibinfo{pages}{016304} (\bibinfo{year}{2006}).

\bibitem[{\citenamefont{Mueschke}(2008)}]{Mue08}
\bibinfo{author}{\bibfnamefont{N.~J.} \bibnamefont{Mueschke}}, Ph.D. thesis,
  \bibinfo{school}{Texas A and M University} (\bibinfo{year}{2008}).

\bibitem[{\citenamefont{Dimonte et~al.}(2004)\citenamefont{Dimonte, Youngs,
  Dimits, Weber, Marinak, Wunsch, Garsi, Robinson, Andrews, Ramaprabhu
  et~al.}}]{DimYou04}
\bibinfo{author}{\bibfnamefont{G.}~\bibnamefont{Dimonte}},
  \bibinfo{author}{\bibfnamefont{D.~L.} \bibnamefont{Youngs}},
  \bibinfo{author}{\bibfnamefont{A.}~\bibnamefont{Dimits}},
  \bibinfo{author}{\bibfnamefont{S.}~\bibnamefont{Weber}},
  \bibinfo{author}{\bibfnamefont{M.}~\bibnamefont{Marinak}},
  \bibinfo{author}{\bibfnamefont{S.}~\bibnamefont{Wunsch}},
  \bibinfo{author}{\bibfnamefont{C.}~\bibnamefont{Garsi}},
  \bibinfo{author}{\bibfnamefont{A.}~\bibnamefont{Robinson}},
  \bibinfo{author}{\bibfnamefont{M.}~\bibnamefont{Andrews}},
  \bibinfo{author}{\bibfnamefont{P.}~\bibnamefont{Ramaprabhu}},
  \bibnamefont{et~al.}, \bibinfo{journal}{Phys. Fluids}
  \textbf{\bibinfo{volume}{16}}, \bibinfo{pages}{1668} (\bibinfo{year}{2004}).

\bibitem[{\citenamefont{Youngs}(2003)}]{You03}
\bibinfo{author}{\bibfnamefont{D.~L.} \bibnamefont{Youngs}},
  \bibinfo{type}{Tech. Rep.} \bibinfo{number}{4102},
  \bibinfo{institution}{American Institute of Aeronautics and Astronautics}
  (\bibinfo{year}{2003}), \bibinfo{note}{presented at the 16th AIAA
  Computational Fluid Dynamics Conference}.

\bibitem[{\citenamefont{Glimm et~al.}(2013)\citenamefont{Glimm, Sharp, Kaman,
  and Lim}}]{GliShaKam11}
\bibinfo{author}{\bibfnamefont{J.}~\bibnamefont{Glimm}},
  \bibinfo{author}{\bibfnamefont{D.~H.} \bibnamefont{Sharp}},
  \bibinfo{author}{\bibfnamefont{T.}~\bibnamefont{Kaman}}, \bibnamefont{and}
  \bibinfo{author}{\bibfnamefont{H.}~\bibnamefont{Lim}},
  \bibinfo{journal}{Phil. Trans. R. Soc. A} \textbf{\bibinfo{volume}{371}},
  \bibinfo{pages}{20120183} (\bibinfo{year}{2013}), \bibinfo{note}{{L}os Alamos
  National Laboratory Preprint LA-UR 11-00423 and Stony Brook University
  Preprint SUNYSB-AMS-11-01}.

\bibitem[{\citenamefont{Kolmogorov}(1962)}]{Kol62}
\bibinfo{author}{\bibfnamefont{A.~N.} \bibnamefont{Kolmogorov}},
  \bibinfo{journal}{J. Fluid Mechanics} \textbf{\bibinfo{volume}{13}},
  \bibinfo{pages}{82} (\bibinfo{year}{1962}).

\bibitem[{\citenamefont{She and Zhang}(2009)}]{SheZha09}
\bibinfo{author}{\bibfnamefont{Z.~S.} \bibnamefont{She}} \bibnamefont{and}
  \bibinfo{author}{\bibfnamefont{Z.~X.} \bibnamefont{Zhang}},
  \bibinfo{journal}{Acta Mech. Sin.} \textbf{\bibinfo{volume}{25}},
  \bibinfo{pages}{279} (\bibinfo{year}{2009}).

\bibitem[{\citenamefont{Liu and She}(2003)}]{LiuShe03}
\bibinfo{author}{\bibfnamefont{L.}~\bibnamefont{Liu}} \bibnamefont{and}
  \bibinfo{author}{\bibfnamefont{Z.-S.} \bibnamefont{She}},
  \bibinfo{journal}{Fluid Dynamics Res.} \textbf{\bibinfo{volume}{33}},
  \bibinfo{pages}{261} (\bibinfo{year}{2003}).

\bibitem[{\citenamefont{Zingale et~al.}(2017)\citenamefont{Zingale, Almgren,
  Sazo, Beckner, Bell, Friesen, Jacobs, Katz, Malone, Nonaka
  et~al.}}]{ZinAlmBar17}
\bibinfo{author}{\bibfnamefont{M.}~\bibnamefont{Zingale}},
  \bibinfo{author}{\bibfnamefont{A.~S.} \bibnamefont{Almgren}},
  \bibinfo{author}{\bibfnamefont{M.~G.~B.} \bibnamefont{Sazo}},
  \bibinfo{author}{\bibfnamefont{V.~E.} \bibnamefont{Beckner}},
  \bibinfo{author}{\bibfnamefont{J.~B.} \bibnamefont{Bell}},
  \bibinfo{author}{\bibfnamefont{B.}~\bibnamefont{Friesen}},
  \bibinfo{author}{\bibfnamefont{A.~M.} \bibnamefont{Jacobs}},
  \bibinfo{author}{\bibfnamefont{M.}~\bibnamefont{Katz}},
  \bibinfo{author}{\bibfnamefont{C.~M.} \bibnamefont{Malone}},
  \bibinfo{author}{\bibfnamefont{A.~J.} \bibnamefont{Nonaka}},
  \bibnamefont{et~al.}, \bibinfo{journal}{ArXive: 1771-06203}
  (\bibinfo{year}{2017}).

\bibitem[{\citenamefont{Calder et~al.}(2012)\citenamefont{Calder, Krueger,
  Jackson, Townsley, Brown, and Times}}]{CalKruJac12}
\bibinfo{author}{\bibfnamefont{A.}~\bibnamefont{Calder}},
  \bibinfo{author}{\bibfnamefont{B.}~\bibnamefont{Krueger}},
  \bibinfo{author}{\bibfnamefont{A.}~\bibnamefont{Jackson}},
  \bibinfo{author}{\bibfnamefont{D.}~\bibnamefont{Townsley}},
  \bibinfo{author}{\bibfnamefont{E.}~\bibnamefont{Brown}}, \bibnamefont{and}
  \bibinfo{author}{\bibfnamefont{F.}~\bibnamefont{Times}},
  \bibinfo{journal}{arXive: 1205-0966}  (\bibinfo{year}{2012}).

\bibitem[{\citenamefont{Lee}(2008)}]{Lee08}
\bibinfo{author}{\bibfnamefont{J.}~\bibnamefont{Lee}},
  \emph{\bibinfo{title}{The Detonation Phenomena}}
  (\bibinfo{publisher}{Cambridge University Press}, \bibinfo{year}{2008}).

\bibitem[{\citenamefont{Glimm}(2018)}]{Gli18}
\bibinfo{author}{\bibfnamefont{J.}~\bibnamefont{Glimm}},
  \bibinfo{journal}{arXive:1205-0966}  (\bibinfo{year}{2018}),
  \bibinfo{note}{1806 06054}.

\bibitem[{\citenamefont{Melvin et~al.}(2015)\citenamefont{Melvin, Lim, Rana,
  Cheng, Glimm, Sharp, and Wilson}}]{MelLimRan14}
\bibinfo{author}{\bibfnamefont{J.}~\bibnamefont{Melvin}},
  \bibinfo{author}{\bibfnamefont{H.}~\bibnamefont{Lim}},
  \bibinfo{author}{\bibfnamefont{V.}~\bibnamefont{Rana}},
  \bibinfo{author}{\bibfnamefont{B.}~\bibnamefont{Cheng}},
  \bibinfo{author}{\bibfnamefont{J.}~\bibnamefont{Glimm}},
  \bibinfo{author}{\bibfnamefont{D.~H.} \bibnamefont{Sharp}}, \bibnamefont{and}
  \bibinfo{author}{\bibfnamefont{D.~C.} \bibnamefont{Wilson}},
  \bibinfo{journal}{Physics of Plasmas} \textbf{\bibinfo{volume}{22}},
  \bibinfo{pages}{022708} (\bibinfo{year}{2015}).

\end{thebibliography}
\end{document}